\documentclass[12pt]{article}
\usepackage{amssymb}

\begin{document}

\begin{titlepage}
\title{\bf Complex Dynamics Effect on Distributions}
\author{ Mehmet Tekkoyun \footnote{E-mail address: tekkoyun@pau.edu.tr; Tel: +902582953616; Fax: +902582953593}\\
{\small Department of Mathematics, Pamukkale University,}\\
{\small 20070 Denizli, Turkey}}
\date{\today}
\maketitle
\begin{abstract}
In this study, Lagrangian and Hamiltonian systems, which are
mathematical models of mechanical systems, were introduced on the
horizontal and the vertical distributions of tangent and cotangent
bundles. Finally, some geometrical and physical results related
to  Lagrangian and Hamiltonian dynamical systems were deduced.
\end{abstract}
\end{titlepage}

\section{Introduction}

In modern differential geometry, the tangent-cotangent bundles of any
differentiable manifold $M$ are seen to be phase-spaces of velocity and
momentum of a given configuration space. Therefore Lagrangian and
Hamiltonian systems in Classical Mechanics are explained by means of basic
structures on the bundles so that this structures can be given by Liouville
vector field $C$, Liouville form $\lambda $, tangent structure $J$, complex
structure $F$, vertical distribution $V$, horizontal distribution $H$ and
semispray $X,$ . The following symbolical equation expresses the dynamical
equations for both Lagrangian and Hamiltonian systems:
\begin{equation}
i_{X}\Phi =\digamma  \label{1.1}
\end{equation}
If anybody studies the Lagrangian systems then equation (\ref{1.1}) is the
intrinsical form of Euler-Lagrange equations, where $\Phi =\Phi
_{L}=-dd_{F}L $ and $\digamma =dE_{L}$ such that $L$ defined by $%
L:TM\rightarrow \mathbf{R} $ is Lagrangian function, $E_{L}\mathbf{\ }$given
by $E_{L}=C(L)-L$ is energy function associated to $L.$

If one studies the Hamiltonian theory then equation (\ref{1.1}) is the
intrinsical form of Hamiltonian equations, where $\Phi =\phi _{\mathbf{H}%
}=-d\lambda $ , $\digamma =d\mathbf{H}$ and $\mathbf{H}$ is a $C^{\infty }-$
function such that $\mathbf{H}:T^{*}M\rightarrow \mathbf{R.}$ Mathematical
expressions of mechanical systems are generally given by the Hamiltonian and
Lagrangian systems. These expressions, in particular geometric expressions
in mechanics and dynamics, are given in some studies \cite{mcrampin,
nutku,deleon,deleon1}. It is known that Lagrangian distribution on
symplectic manifolds is used in geometric quantization and a connection on a
symplectic manifold is an important structure to obtain a deformation
quantization \cite{etayo}. Paracomplex geometry and the framework of
para-K\"{a}hlerian manifolds being geometric model of generalized Lagrange
spaces were introduced \cite{cruceanu}. Para-complex analogues of the
Lagrangians and Hamiltonians were obtained in the framework of
para-K\"{a}hlerian manifold and the geometric conclusions on a para-complex
dynamical systems were obtained \cite{tekkoyun2005}. As determined
references and there in, real-complex-paracomplex geometry and
mechanical-dynamical systems were analyzed successfully, but they have not
dealt with complex dynamical systems on horizontal distribution and vertical
distribution of tangent-cotangent bundles of any manifold $M$.

Therefore, here, Euler-Lagrange and Hamiltonian equations related to complex
dynamical systems on the distributions used in obtaining geometric
quantization have been obtained.

\section{Preliminaries}

In this study all mappings are considered of the class $C^{\infty },$\
expressed by the words ''differentiate'' or ''smooth''. \ The indices $i$,$j$%
,...run over set $\{1,..,n\}$\ and Einstein convention of summarizing is
adopted over all this paper. $\mathbf{R}$, $\mathcal{F}(TM)$, $\chi (TM)$
and $\chi (T^{*}M)$ denote the set of real numbers, the set of real
functions on $TM$, the set of vector fields on $TM$ and the set of 1-forms
on $T^{*}M,$ respectively.

\subsection{Basic Structures}

In this subsection, some definitions were derived taking from \cite{radu}.
Let $TM$ be tangent bundle of a real $n$-dimensional differentiable manifold
$M$. Then it will be denoted a point of $M$ by $x$ and its local coordinate
system by $(U,\varphi )$ such that $\varphi (x)=(x^{i}).$\ Such that the
projection $\pi :TM\rightarrow M$, $\pi (u)=x,$\ a point $u\in TM$ will be
denoted by $(x,y)$, its local coordinates being $(x^{i},y^{i})$. There are
the natural basis $(\frac{\partial }{\partial x^{i}},\frac{\partial }{%
\partial y^{i}})$ and dual basis $(dx^{i},dy^{i})$\ of the tangent space $%
T_{u}TM$\ and the cotangent space $T_{u}^{*}(TM)$ at the point $u\in TM$%
\textbf{, }respectively$.$\ Consider the $\mathcal{F}(TM)-$ and $\mathcal{F}%
(T^{*}M)-$ linear mappings $J:\chi (TM)\rightarrow \chi (TM)$\thinspace and $%
J^{*}:\chi (T^{*}M)\rightarrow \chi (T^{*}M)$ given by
\[
\begin{array}{l}
J(\frac{\partial }{\partial x^{i}})=\frac{\partial }{\partial y^{i}},\,J(%
\frac{\partial }{\partial y^{i}})=0,\,
\end{array}
\]
and
\[
\begin{array}{l}
J^{*}(dx^{i})=dy^{i},J^{*}(dy^{i})=0.
\end{array}
\]
The tangent space $V_{u}$ to the fibre $\pi ^{-1}(x)$\ in the point $u\in TM$%
\ is locally spanned by $\{\frac{\partial }{\partial y^{1}},..,\frac{%
\partial }{\partial y^{n}}\}$. Therefore, the mapping $V$: $u\in
TM\rightarrow V_{u}\subset T_{u}TM$\ provides a regular distribution
generated by the adapted basis $\{\frac{\partial }{\partial y^{i}}\}.$\
Consequently, $V$\ is an integrable distribution on $TM$. $V$\ is called the
vertical distribution on $TM$.\ Let $N$ be a nonlinear connection on $TM$. $%
N $ is characterized by $v$, $h$ vertical and horizontal projectors. We
consider the vertical projector $v:\chi (TM)\rightarrow \chi (TM)$ defined
by $v(X)=X,\,\forall \,X\in \chi (VTM);$ $v(X)=0,\forall \,X\in \chi (HTM).$
Similarly, the mapping $H$: $u\in TM\rightarrow H_{u}\subset T_{u}TM$\
provides a regular distribution determined by the adapted basis $\{\frac{%
\delta }{\delta x^{i}}\}.$\ Consequently, $H$\ is an integrable distribution
on $TM$. $H$\ is called the horizontal distribution on $TM$.\ There is a $%
\mathcal{F}(TM)-$linear mapping $h:\chi (TM)\rightarrow \chi (TM),$ for
which $h^{2}=h,$ $Ker$ $h=\chi (VTM).$ Any vector field $X\in \chi (TM)$ can
be uniquely written as follows $X=hX+vX=X^{H}+X^{V}$. Therefore $X^{H}$ and $%
X^{V}$ are horizontal and vertical components of vector field $X$.
Therefore, any vector field $X$ can be uniquely written in the form
\[
\begin{array}{l}
X=X^{H}+X^{V}
\end{array}
\]
such that
\[
\begin{array}{l}
X^{H}=X^{i}(\frac{\partial }{\partial x^{i}}-N_{j}^{i}(x,y)\frac{\partial }{%
\partial y^{j}}),\,\,\,X^{V}=X^{i}N_{j}^{i}(x,y)\frac{\partial }{\partial
y^{j}}
\end{array}
\]
where $N_{j}^{i}$ is local coefficient of nonlinear connection $N$ on $TM.$

$(\frac{\delta }{\delta x^{i}},\frac{\partial }{\partial y^{i}})$ is a local
basis adapted to the horizontal distribution $HTM$ and the vertical
distribution $VTM$. Then $(dx^{i},\delta y^{i})$ is dual basis of $(\frac{%
\delta }{\delta x^{i}},\frac{\partial }{\partial y^{i}})$ basis. We have
\[
\begin{array}{l}
\frac{\delta }{\delta x^{i}}=\frac{\partial }{\partial x^{i}}-N_{j}^{i}(x,y)%
\frac{\partial }{\partial y^{j}}.
\end{array}
\]
and
\[
\begin{array}{l}
\delta y^{i}=dy^{i}+N_{j}^{i}(x,y)dx^{j}.
\end{array}
\]
$F$ is an almost complex structure on $TM.$ $F^{*}$ is the dual structure of
$F.$ For the operators $h,v,F,F^{*}$ we get
\[
\begin{array}{l}
h+v=I;\,\,F^{2}=-I;\,F^{*2}=-I \\
h(\frac{\delta }{\delta x^{i}})=\frac{\delta }{\delta x^{i}};\,h(\frac{%
\partial }{\partial y^{i}})=0;\,v(\frac{\delta }{\delta x^{i}})=0;v(\frac{%
\partial }{\partial y^{i}})=\frac{\partial }{\partial y^{i}},\, \\
F(\frac{\delta }{\delta x^{i}})=-\frac{\partial }{\partial y^{i}};\,\,F(%
\frac{\partial }{\partial y^{i}})=\frac{\delta }{\delta x^{i}}, \\
F^{*}(dx^{i})=-\delta y^{i};\,F^{*}(\delta y^{i})=dx^{i}.
\end{array}
\]
\textbf{Lagrangian Dynamical Systems}

In this section, Euler-Lagrange equations for Classical Mechanics are made
by means of almost complex structure $F$\ under the consideration of the
basis $\{\frac{\delta }{\delta x^{i}},\,\frac{\partial }{\partial y^{i}}\}$
on distributions $HTM$ and $VTM$ of tangent bundle $TM$ of manifold $M.$ Let
$(x^{i},y^{i})$ be its local coordinates. Let semispray be the vector field $%
X$ given by
\begin{equation}
\begin{array}{l}
X=X^{i}\frac{\delta }{\delta x^{i}}+\stackrel{.}{X}^{i}\frac{\partial }{%
\partial y^{i}},\stackrel{.}{\,\,X}^{i}=X^{i}N_{j}^{i}
\end{array}
\label{3.1}
\end{equation}
where the dot indicates the derivative with respect to time $t$. The vector
field denoted by $C=F(X)$ and expressed by
\begin{equation}
\begin{array}{l}
C=-X^{i}\frac{\partial }{\partial y^{i}}\stackrel{.}{+X}^{i}\frac{\delta }{%
\delta x^{i}}
\end{array}
\label{3.2}
\end{equation}
is called \textit{Liouville vector field} on the bundle $TM$. The maps given
by $\mathbf{T,P}:TM\rightarrow \mathbf{R}$ such that $\mathbf{T}=\frac{1}{2}%
m_{i}(x^{i})^{2},\mathbf{P}=m_{i}gh$ are called \textit{the kinetic energy}
and \textit{the potential energy of the mechanical system,} respectively.%
\textit{\ }Here\textit{\ }$m_{i},g$ and $h$ stand for mass of a mechanical
system having $m$ particles, the gravity acceleration and distance to the
origin of a mechanical system on the tangent bundle $TM$, respectively. Then
$L:TM\rightarrow \mathbf{R}$ is a map that satisfies the conditions; i) $L=%
\mathbf{T-P}$ is a \textit{Lagrangian function, ii)} the function given by $%
E_{L}=C(L)-L$ is \textit{a Lagrangian energy}. The operator $i_{F}$ induced
by $F$ and shown by
\begin{equation}
\begin{array}{l}
i_{F}\omega (X_{1},X_{2},...,X_{r})=\sum_{i=1}^{r}\omega
(X_{1},...,F(X_{i}),...,X_{r})
\end{array}
\label{3.3}
\end{equation}
is said to be \textit{vertical derivation, }where $\omega \in \wedge
^{r}{}TM,$ $X_{i}\in \chi (TM).$ The \textit{vertical differentiation} $d_{F}
$ is defined by
\begin{equation}
\begin{array}{l}
d_{F}=[i_{F},d]=i_{F}d-di_{F}
\end{array}
\label{3.4}
\end{equation}
where $d$ is the usual exterior derivation. For an almost complex structure $%
F$, the closed fundamental form is the closed 2-form given by $\Phi
_{L}=-dd_{F}L$ such that
\begin{equation}
\begin{array}{l}
d_{F}:\mathcal{F}(TM)\rightarrow {}T^{*}M
\end{array}
\label{3.5}
\end{equation}
Then we have
\begin{equation}
\begin{array}{l}
\Phi _{L}=-(\frac{\delta }{\delta x^{j}}dx^{j}+\frac{\partial }{\partial
y^{j}}\delta y^{j})(-\frac{\partial L}{\partial y^{i}}dx^{i}+\frac{\delta L}{%
\delta x^{i}}\delta y^{i}) \\
\,\,\,\,\,\,\,\,\,\,=\frac{\delta (\partial L)}{\delta x^{j}\partial y^{i}}%
dx^{j}\wedge dx^{i}-\frac{\delta ^{2}L}{\delta x^{j}\delta x^{i}}%
dx^{j}\wedge \delta y^{i}+\frac{\partial ^{2}L}{\partial y^{j}\partial y^{i}}%
\delta y^{j}\wedge dx^{i}-\frac{\partial (\delta L)}{\partial y^{j}\delta
x^{i}}\delta y^{j}\wedge \delta y^{i}.
\end{array}
\label{3.6}
\end{equation}
Let $X$ be the second order differential equation (semispray) determined by
\textbf{Eq. }(\ref{1.1}). Then we get
\begin{equation}
\begin{array}{l}
i_{X}\Phi _{L}=\Phi _{L}(X)=X^{i}\frac{\delta (\partial L)}{\delta
x^{j}\partial y^{i}}\delta _{i}^{j}dx^{i}-X^{i}\frac{\delta (\partial L)}{%
\delta x^{j}\partial y^{i}}dx^{j}-X^{i}\frac{\delta ^{2}L}{\delta
x^{j}\delta x^{i}}\delta _{i}^{j}\delta y^{i}+\stackrel{.}{X}^{i}\frac{%
\delta ^{2}L}{\delta x^{j}\delta x^{i}}dx^{j} \\
+\stackrel{.}{X}^{i}\frac{\partial ^{2}L}{\partial y^{j}\partial y^{i}}%
\delta _{i}^{j}dx^{i}-X^{i}\frac{\partial ^{2}L}{\partial y^{j}\partial y^{i}%
}\delta y^{j}-\stackrel{.}{X}^{i}\frac{\partial (\delta L)}{\partial
y^{j}\delta x^{i}}\delta _{i}^{j}\delta y^{i}+\stackrel{.}{X}^{i}\frac{%
\partial (\delta L)}{\partial y^{j}\delta x^{i}}\delta y^{j}.
\end{array}
\label{3.7}
\end{equation}
Since the closed 2-form $\Phi _{L}$ on $TM$ is the symplectic structure, it
is found
\begin{equation}
\begin{array}{l}
E_{L}=C(L)-L=-X^{i}\frac{\partial L}{\partial y^{i}}+\stackrel{.}{X}^{i}%
\frac{\delta L}{\delta x^{i}}-L
\end{array}
\label{3.8}
\end{equation}
and hence
\begin{equation}
\begin{array}{l}
dE_{L}=-X^{i}\frac{\delta (\partial L)}{\delta x^{j}\partial y^{i}}dx^{j}+%
\stackrel{.}{X}^{i}\frac{\delta ^{2}L}{\delta x^{j}\delta x^{i}}dx^{j}-\frac{%
\delta L}{\delta x^{j}}dx^{j}-X^{i}\frac{\partial ^{2}L}{\partial
y^{j}\partial y^{i}}\delta y^{j}+\stackrel{.}{X}^{i}\frac{\partial (\delta L)%
}{\partial y^{j}\delta x^{i}}\delta y^{j}-\frac{\partial L}{\partial y^{j}}%
\delta y^{j}
\end{array}
\label{3.9}
\end{equation}
With the use of \textbf{Eq.} (\ref{1.1}), considering \textbf{Eqs.}(\ref{3.7}%
) and (\ref{3.9}) we get
\begin{equation}
\begin{array}{l}
X^{i}\frac{\delta (\partial L)}{\delta x^{j}\partial y^{i}}\delta
_{i}^{j}dx^{i}-X^{i}\frac{\delta (\partial L)}{\delta x^{j}\partial y^{i}}%
dx^{j}-X^{i}\frac{\delta ^{2}L}{\delta x^{j}\delta x^{i}}\delta
_{i}^{j}\delta y^{i}+\stackrel{.}{X}^{i}\frac{\delta ^{2}L}{\delta
x^{j}\delta x^{i}}dx^{j} \\
+\stackrel{.}{X}^{i}\frac{\partial ^{2}L}{\partial y^{j}\partial y^{i}}%
\delta _{i}^{j}dx^{i}-X^{i}\frac{\partial ^{2}L}{\partial y^{j}\partial y^{i}%
}\delta y^{j}-\stackrel{.}{X}^{i}\frac{\partial (\delta L)}{\partial
y^{j}\delta x^{i}}\delta _{i}^{j}\delta y^{i}+\stackrel{.}{X}^{i}\frac{%
\partial (\delta L)}{\partial y^{j}\delta x^{i}}\delta y^{j} \\
=-X^{i}\frac{\delta (\partial L)}{\delta x^{j}\partial y^{i}}dx^{j}+%
\stackrel{.}{X}^{i}\frac{\delta ^{2}L}{\delta x^{j}\delta x^{i}}dx^{j}-\frac{%
\delta L}{\delta x^{j}}dx^{j}-X^{i}\frac{\partial ^{2}L}{\partial
y^{j}\partial y^{i}}\delta y^{j}+\stackrel{.}{X}^{i}\frac{\partial (\delta L)%
}{\partial y^{j}\delta x^{i}}\delta y^{j}-\frac{\partial L}{\partial y^{j}}%
\delta y^{j}
\end{array}
\label{3.10}
\end{equation}
or
\begin{equation}
\begin{array}{l}
X^{i}\frac{\delta (\partial L)}{\delta x^{j}\partial y^{i}}dx^{j}-X^{i}\frac{%
\delta ^{2}L}{\delta x^{j}\delta x^{i}}\delta y^{j}+\stackrel{.}{X}^{i}\frac{%
\partial ^{2}L}{\partial y^{j}\partial y^{i}}dx^{j}-\stackrel{.}{X}^{i}\frac{%
\partial (\delta L)}{\partial y^{j}\delta x^{i}}\delta y^{j}+\frac{\delta L}{%
\delta x^{j}}dx^{j}+\frac{\partial L}{\partial y^{j}}\delta y^{j}=0
\end{array}
\label{3.11}
\end{equation}
If a curve denoted by $\alpha :\mathbf{R}\rightarrow TM$ is considered to be
an integral curve of $X,$ i.e$.$ $X(\alpha (t))=\frac{d\alpha (t)}{dt}$ then
we have the equations
\begin{equation}
\begin{array}{l}
\frac{d}{dt}(\frac{\partial L}{\partial y^{i}})+\frac{\delta L}{\delta x^{i}}%
=0,\,\,\frac{d}{dt}(\frac{\delta L}{\delta x^{i}})-\frac{\partial L}{%
\partial y^{i}}=0,
\end{array}
\label{3.12}
\end{equation}
which are named to be a \textit{Euler-Lagrange equations} which are deduced
by means of an almost complex structure $F$ on $HTM$ horizontal and $VTM$
vertical distributions.

Thus the triple $(TM,\Phi _{L},X)$ is a \textit{mechanical system }which is%
\textit{\ }structured by means of an almost complex structure $F$ and taking
into account the basis $\{\frac{\delta }{\delta x^{i}},\frac{\partial }{%
\partial y^{i}}\}$ on the distributions $HTM$ and $VTM$.

\section{Hamiltonian Dynamical Systems}

In this section, Hamiltonian equations for Classical Mechanics are obtained
on the distributions ${}HT^{*}M$ and $VT^{*}M$ of $T^{*}M.$ Suppose that an
almost complex structure, a Liouville form and a 1-form on $T^{*}M$ are
shown by $P^{*}$, $\lambda $ and $\omega $, respectively$.$ Then we have
\begin{equation}
\begin{array}{l}
\omega =\frac{1}{2}(x^{i}dx^{i}+y^{i}\delta y^{i})
\end{array}
\label{4.1}
\end{equation}
and
\begin{equation}
\begin{array}{l}
\lambda =P^{*}(\omega )=\frac{1}{2}(-x^{i}\delta y^{i}+y^{i}dx^{i}).
\end{array}
\end{equation}
It is known that if $\phi $ is a closed 2- form on $T^{*}M,$ then $\phi _{%
\mathbf{H}}$ is also a symplectic structure on ${}{}T^{*}M$. If Hamiltonian
vector field $X_{\mathbf{H}}$ associated with Hamiltonian energy $\mathbf{H}$
is given by
\begin{equation}
\begin{array}{l}
X_{\mathbf{H}}=X^{i}\frac{\delta }{\delta x^{i}}+Y^{i}\frac{\partial }{%
\partial y^{i}},
\end{array}
\label{4.3}
\end{equation}
then we deduce
\begin{equation}
\begin{array}{l}
\phi _{\mathbf{H}}=-d\lambda =\delta y^{i}\wedge dx^{i}
\end{array}
\end{equation}
and
\begin{equation}
\begin{array}{l}
i_{X_{\mathbf{H}}}\phi =Y^{i}dx^{i}-X^{i}\delta y^{i}.
\end{array}
\label{4.5}
\end{equation}
Moreover, the differential of Hamiltonian energy is written as follows:
\begin{equation}
\begin{array}{l}
d\mathbf{H}=\frac{\delta \mathbf{H}}{\delta x^{i}}dx^{i}+\frac{\partial
\mathbf{H}}{\partial y^{i}}\delta y^{i}.
\end{array}
\label{4.6}
\end{equation}
By means of \textbf{Eq.}(\ref{1.1}), using \textbf{Eqs. }(\ref{4.5}) and (%
\ref{4.6}), the Hamiltonian vector field is calculated to be
\begin{equation}
\begin{array}{l}
X_{\mathbf{H}}=-\frac{\partial \mathbf{H}}{\partial y^{i}}\frac{\delta }{%
\delta x^{i}}+\frac{\delta \mathbf{H}}{\delta x^{i}}\frac{\partial }{%
\partial y^{i}}.
\end{array}
\label{4.7}
\end{equation}
Suppose that a curve
\begin{equation}
\begin{array}{l}
\alpha {:\,}I\subset \mathbf{R}\rightarrow T^{*}M
\end{array}
\label{4.8}
\end{equation}
is an integral curve of the Hamiltonian vector field $X_{\mathbf{H}}$, i.e.,
\begin{equation}
\begin{array}{l}
X_{\mathbf{H}}(\alpha (t))=\frac{d\alpha (t)}{dt},\,\,t\in I.
\end{array}
\label{4.9}
\end{equation}
In the local coordinates, if it is considered to be
\begin{equation}
\begin{array}{l}
\alpha (t)=(x^{i}(t),y^{i}(t)),
\end{array}
\label{4.10}
\end{equation}
we obtain
\begin{equation}
\begin{array}{l}
\frac{d\alpha (t)}{dt}=\frac{dx^{i}}{dt}\frac{\delta }{\delta x^{i}}+\frac{%
dy^{i}}{dt}\frac{\partial }{\partial y^{i}}.
\end{array}
\label{4.11}
\end{equation}
Taking into consideration \textbf{Eqs. }(\ref{4.8}), (\ref{4.6}) and (\ref
{4.10}), we have the equations
\begin{equation}
\begin{array}{l}
\frac{dx^{i}}{dt}=-\frac{\partial \mathbf{H}}{\partial y^{i}},\frac{dy^{i}}{%
dt}=\frac{\delta \mathbf{H}}{\delta x^{i}},
\end{array}
\label{4.12}
\end{equation}
which are named to be \textit{Hamiltonian equations} which are deduced by
means of an almost complex structure $F^{*}$ on the horizontal distribution $%
{}{}HT^{*}M$ and vertical distribution $VT^{*}M.$

Hence the triple $(T^{*}M,\phi _{\mathbf{H}},X_{\mathbf{H}})$ is shown to be
a \textit{Hamiltonian mechanical system }which\textit{\ }are deduced by
means of an almost complex structure $F^{*}$ and using of basis $\{\frac{%
\delta }{\delta x^{i}},\frac{\partial }{\partial y^{i}}\}$ on the
distributions ${}HT^{*}M$ and $VT^{*}M$.

\section{Conclusions}

This paper has obtained to exist physical proof of the both mathematical
equality given by $TM=HTM\oplus VTM$ and its dual equality$.$ Lagrangian and
Hamiltonian dynamics have intrinsically been described by means of an almost
complex structure $F^{*}$ being functional on distributions of tangent and
cotangent bundles $TM$ and $T^{*}M$ of manifold $M$. Also, we deduce that
Euler-Lagrange equations given by (\ref{3.12}) turn into Hamiltonian
equations defined by (\ref{4.12}) taking care into equalities $x^{i}=\frac{%
\delta L}{\delta x^{i}},$ $y^{i}=\frac{\partial L}{\partial y^{i}}$ and $%
H=-L,$ and vice versa.

\section{\textbf{Discussions}}

As is well-known, geometry of Lagrangian and Hamiltonian formalisms give a
model for Relativity, Gauge Theory and Electromagnetism in a very natural
blending of the geometrical structures of the space with the characteristics
properties of these physical fields. Therefore we consider that the
equations (\ref{3.12}) and (\ref{4.12}) especially can be used in fields
determined the above of physical.


\begin{thebibliography}{9}
\bibitem{mcrampin}  M. Crampin, J. Phys. A: Math. Gen. 14\textbf{\ }(1981)
2567.

\bibitem{nutku}  N. Nutku, J. Math. Phys. 25\textbf{\ }(1984) 2007.

\bibitem{deleon}  M. De Leon, P.R. Rodrigues, Methods of Differential
Geometry in Analytical Mechanics, North-Holland Mathematics Studies,
vol.152, Elsevier, Amsterdam, 1989.

\bibitem{deleon1}  M. De Leon, P.R. Rodrigues, Diff. Geom. and its Appl.,
Proceedings of the Conference, August 24-30 (1986) 179.

\bibitem{etayo}  F. Etayo, R. Santanar\'{i}a, U.R. Tr\'{i}as, Diff. Geom.
and its Appl., 24 (2006) 33.

\bibitem{cruceanu}  V. Cruceanu, P.M. Gadea, J. M. Masqu\'{e},
Para-Hermitian and Para- K\"{a}hler Manifolds, Supported by the commission
of the European Communities' Action for Cooperation in Sciences and
Technology with Central Eastern European Countries n. ERB3510PL920841.

\bibitem{tekkoyun2005}  M. Tekkoyun, Phys.\ Lett. A, 340 (2005) 7.

\bibitem{radu}  R. Miron, D. Hrimiuc, H. Shimada, S. V. Sabau, The Geometry
of Hamilton and Lagrange Spaces, Kluwer Academic Publishers, 2001.
\end{thebibliography}
\end{document}